\documentclass[aps,twocolumn,PRB,reprint,superscriptaddress]{revtex4-1}
\pdfoutput=1
\usepackage{amsmath}
\usepackage{epsfig}
\usepackage{graphicx}
\usepackage{graphicx, color, epstopdf}
\usepackage{bm}
\usepackage{amssymb}
\usepackage{hyperref}
\usepackage{xcolor}
\usepackage{subfigure}
\begin{document}
\bibliographystyle{apsrev4-1}
\title{Fabry-Perot interference in three dimensional second-order topological insulator constrictions}

\author{Junyu Luo}
\affiliation{College of Physics and Electronic Science, Hubei Normal University, Huangshi, 435002, China}

\author{Kun Luo}
\email{Corresponding author: luokunphys@hbnu.edu.cn}
\affiliation{College of Physics and Electronic Science, Hubei Normal University, Huangshi, 435002, China}

\begin{abstract}
The gapless chiral hinge states of three dimensional second-order topological insulators (SOTIs)
support a quantized conductance plateau on thick nanowire. Here, we numerical study the conductance
of SOTI constrictions. According to finite size effects, the hinge states in narrow region could be hybridized,
which will induce reflection at the two ends of constrictions. The conductance exists the Fabry-Perot
oscillation pattern because of multiple reflections.
We also study the impact of the magnetic field on the Fabry-Perot interference. We show the dimensional
effect that the magnetic field leads to the electrons being localized on two hinges.
Our results are robust against moderate disorder so that we expect these Fabry-Perot patterns could be observed in experiments.
\end{abstract}

\date{\today}

\maketitle

\section{INTRODUCTION}
With the further study of the quantum Hall effect, the notion of topology
is introduced into condensed matter physics, which induce the various
topological materials have been proposed \cite{RevModPhys.82.3045,RevModPhys.83.1057,hasan2011three,moore2010birth}. In these topological
phases, the bulk-edge correspondence is the important feature.
Generally, the number of edge states can be confirmed by the topological
invariant of bulk states. For example, the three dimensional (3D) topological
insulators have surface states which transport along two dimensional surface.
Recently, the different bulk-boundary correspondence has been discovered,
which is called higher-order topological phases \cite{benalcazar2017quantized,PhysRevB.96.245115,PhysRevLett.119.246402,PhysRevLett.119.246401,
schindler2018higher,PhysRevB.98.201114,PhysRevLett.123.216803,PhysRevB.97.205136,PhysRevResearch.2.043223,
PhysRevB.102.094503,PhysRevLett.123.177001,PhysRevB.99.041301,PhysRevB.100.205406,PhysRevLett.121.096803,schindler2018highernp}.
For instance, a 3D second-order topological insulator supports the one dimensional
gapless chiral or helical hinge states instead of 2D surface states.

Simultaneously, the finite size effect of topological materials have
been theoretically and experimentally studied \cite{PhysRevLett.101.246807,PhysRevB.80.205401,PhysRevB.86.245436,takane2016disorder,xiao2015anisotropic,
pan2015electric,PhysRevLett.120.016801,yilmaz2017thickness,collins2018electric,PhysRevB.94.224501,PhysRevB.96.125426,PhysRevB.90.045309}.
The finite size effect in the quantum spin Hall insulator is first studied
by Zhou $et$ $al$  \cite{PhysRevLett.101.246807}, they found that the wavefunctions of helical edge states
are overlapped, leading to an energy gap. Moreover,
Zhou $et$ $al.$ found that the magnetic field can close the energy gap.
One may consider that the finite size
effects in topological material is constructive to application in electronic devices,
such as electronic switch \cite{PhysRevLett.107.086803,PhysRevB.83.081402,PhysRevB.86.165418,liu2011effect,PhysRevLett.123.206801,PhysRevResearch.2.023242,nadeem2022optimizing} and electronic interferometers \cite{PhysRevB.83.165304,PhysRevB.81.235323,takagaki2012backscattering,PhysRevB.103.205124,lu2022topological}.

\begin{figure}
\centering
\includegraphics[width=\columnwidth]{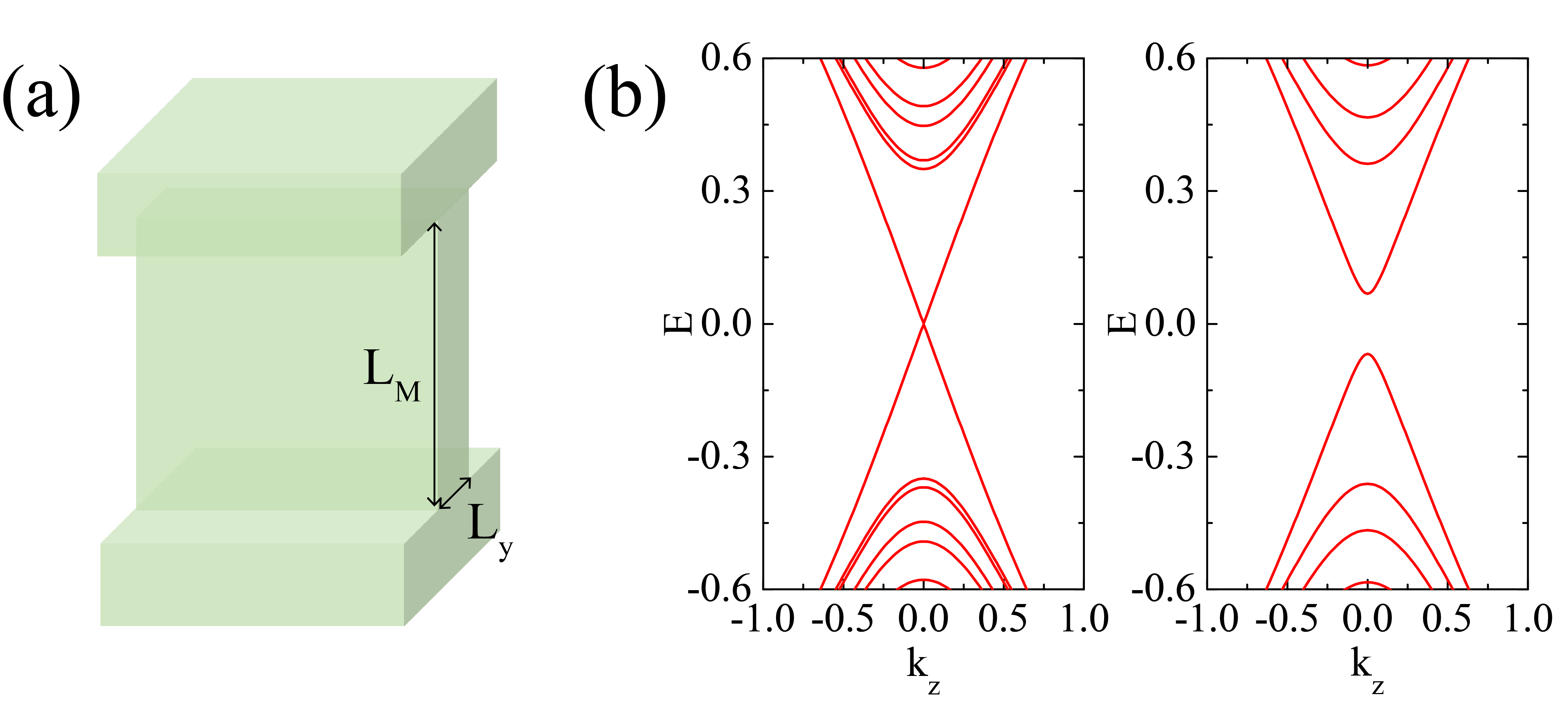}\\
\caption{(a) Sketch of the SOTI constriction with length $L_M$ and width $L_y$.
(b) Energy bands of SOTIs for $L_y=15a$ (left panel) and $L_y=4a$ (right panel).
The model parameters are $t=-1, M=2.3, \Delta_1=1, \Delta_2=0.5$.
}\label{fig1}
\end{figure}

We also find the similar character
of the finite size effect in three dimensional second-order
chiral topological insulators. In this article, we propose an Fabry-Perot interferometer formed by thin quasi-2D
constrictions of 3D SOTIs, see Fig. \ref{fig1} (a).
Due to the finite size effect, the hinge states in the constrictions will
be hybridized together. Thus, there are multiple backscattering at ends of the constrictions.
As a consequence, the differential conductance have an Fabry-Perot
oscillation when potential energy or length of the constrictions are varied.
We also consider the influence of magnetic field, with the emergence of new oscillation pattern
and even the disappearance of zero conductance platform. These numerical results are robust
against moderate disorder because of topologically protected hinge states.

The rest of this paper is organized as follows. In Sec. \ref{md},
we present the model Hamiltonian and discuss the finite size effects.
Detailed numerical calculations on the lattice model is performed in Sec. \ref{rd}.
Finally, a brief summary and outlook are given in Sec. \ref{so}.

\begin{figure}
\centering
\includegraphics[width=\columnwidth]{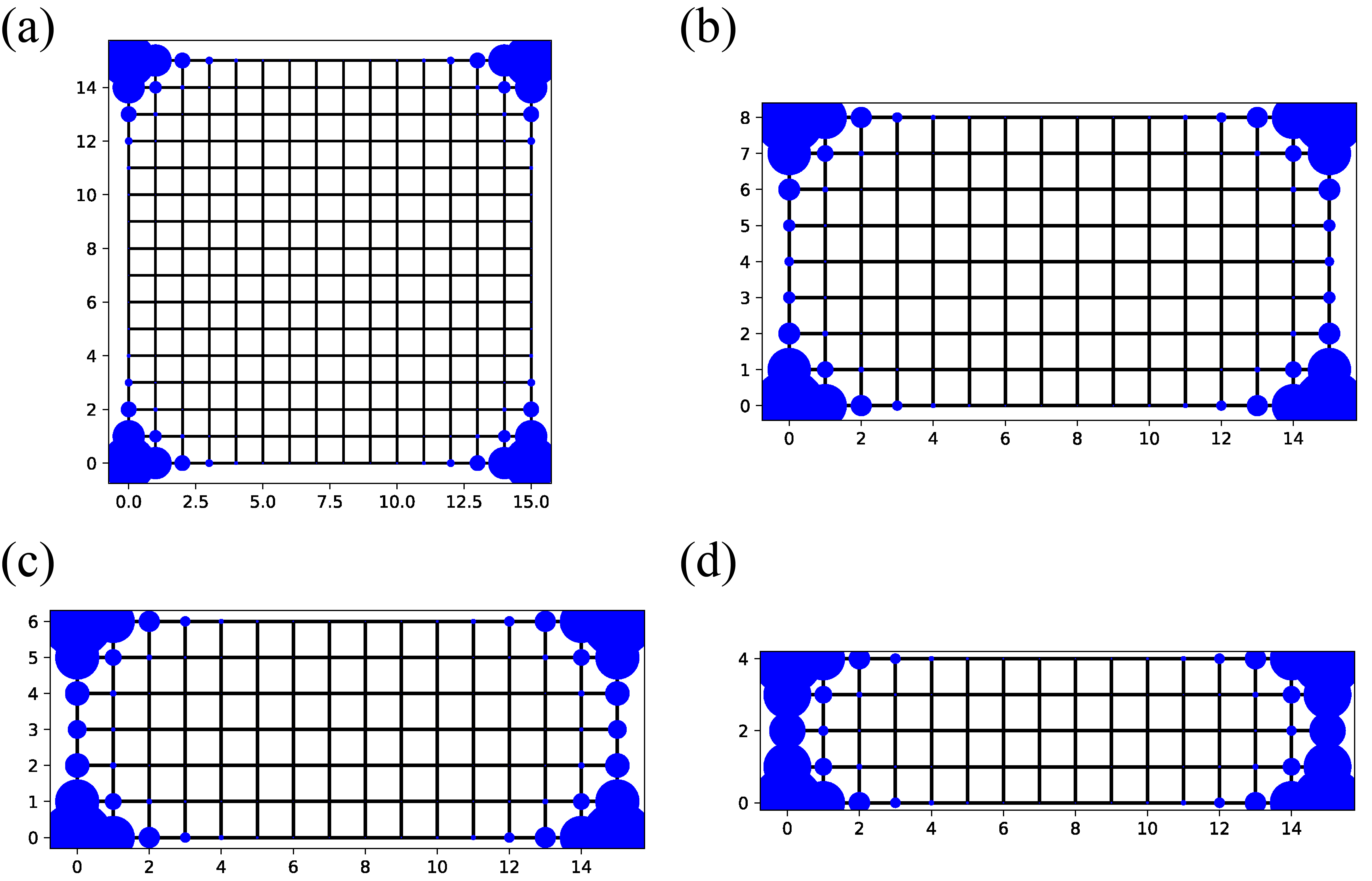}\\
\caption{The spatial distribution of the hinge states with different width $L_y=15a, 8a, 6a, 4a$.
The system parameters are same as Fig. \ref{fig1}.
}\label{fig2}
\end{figure}

\section{Model}\label{md}
We investigate the charge conductance of a 3D SOTIs constrictions,
see Fig. \ref{fig1} (a). Semi-infinite leads of width $W_L$ are
attached to the constrictions. With a small bias voltage,
the electron will propagate along hinge from one lead to another leads.
The Hamiltonian of leads and constrictions is adopted the Schindler's model \cite{schindler2018higher}
\begin{equation}\label{Hk}
\begin{split}
H_{\text{SOTI}}&=\left( M+t\sum_i{\cos k_i} \right) \tau_z\sigma_0 + \Delta_1\sum_i{\sin k_i \tau_x\sigma_i}\\
&\ \ \ +\Delta_2(\cos k_x - \cos k_y)\tau_y\sigma_0,
\end{split}
\end{equation}
where $\sigma_i$ and $\tau_i,i=x,y,z$ are Pauli matrices acting on
the spin and orbital space, respectively. For $1<|M/t|<3$ and $\Delta_1,\Delta_2 \neq 0$,
$H_{\text{SOTI}}$ represents a chiral 3D SOTI.
In Fig. \ref{fig1} (b), we show the energy bands of a nanowire when $x$ and $y$
direction is open condition and $z$ direction is period condition, and the width of
$L_y$ is $15a$ and $4a$, respectively. One can see the gapless bands open a gap
when the confinement is strengthened. The local density of hinge states with different
width $L_y=15a, 8a, 6a, 4a$ is shown in Fig. \ref{fig2}. The hybridized hinge
states which leading to the energy gap are clearly seen.

\section{Results and discussion}\label{rd}

\begin{figure}
\centering
\includegraphics[width=\columnwidth]{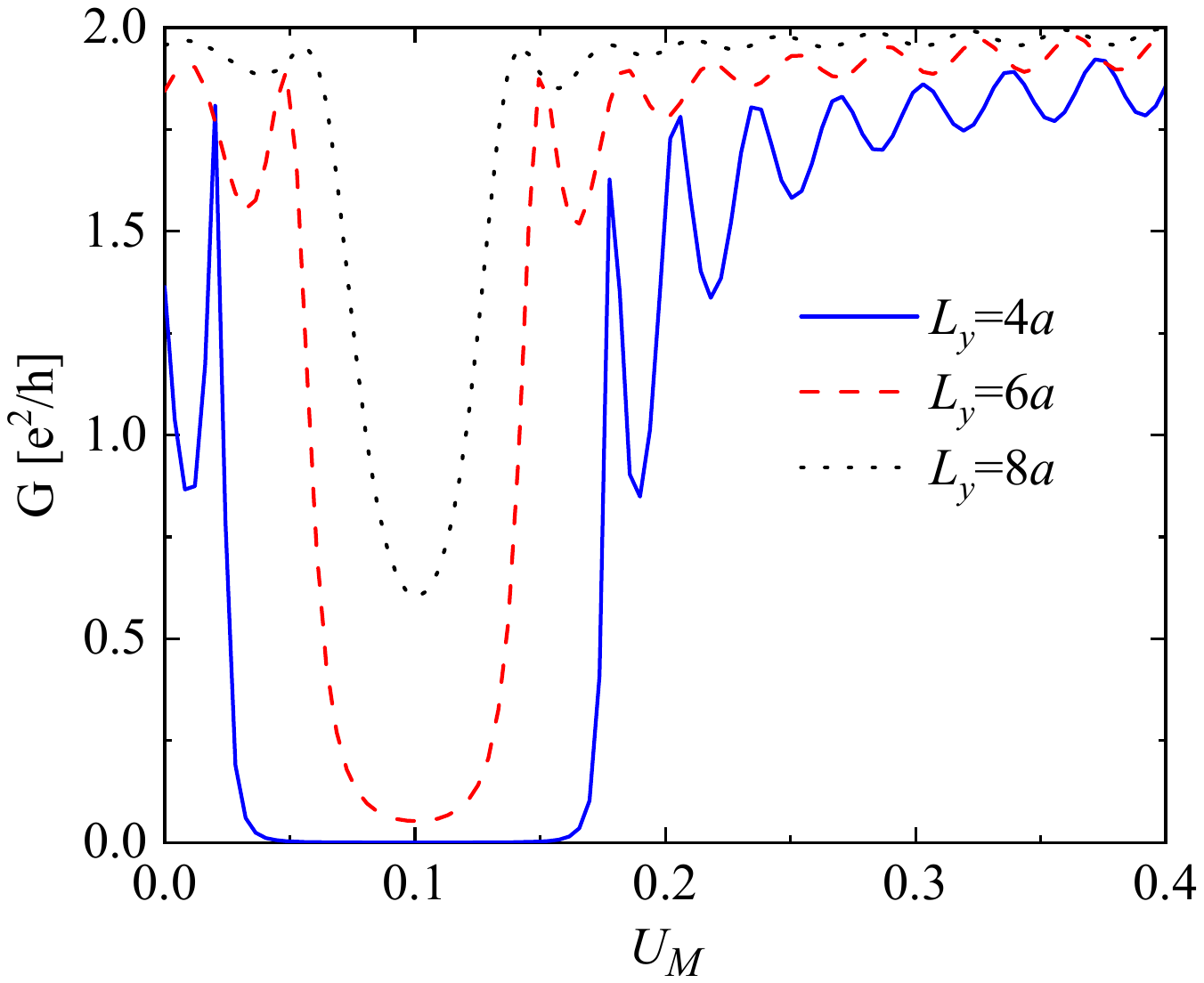}\\
\caption{The transmission probability as a function of potential energy $U_M$
with different thickness $L_y=4a, 6a, 8a$.
The width of leads is $W_L=15a$, the length of constriction is $L_M=80a$.
The system parameters are same as Fig. \ref{fig1}.
}\label{fig3}
\end{figure}

\begin{figure}
\centering
\includegraphics[width=\columnwidth]{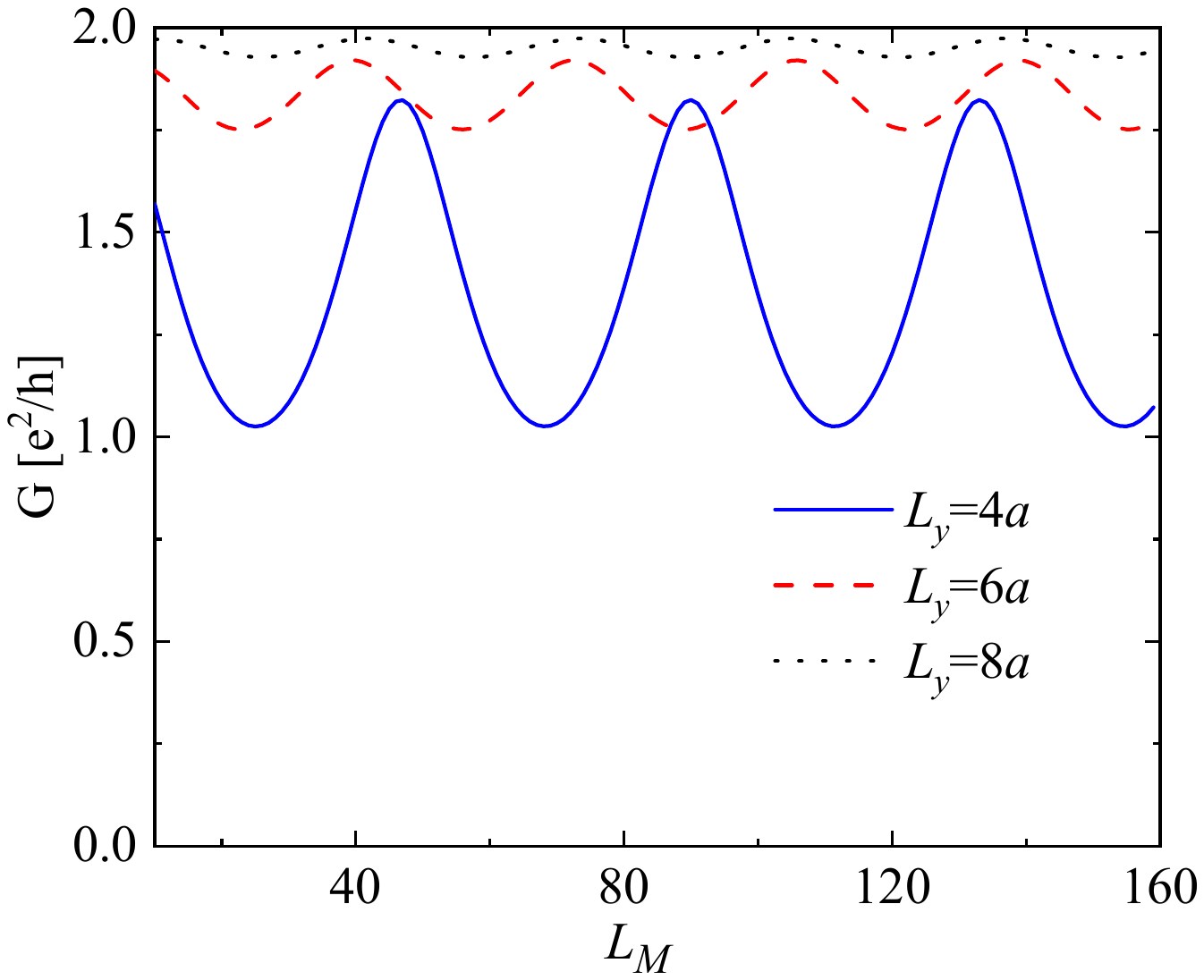}\\
\caption{The transmission probability as a function of constriction length $L_M$
with different thickness $L_y=4a, 6a, 8a$.
The width of leads is $W_L=15a$, the potential energy is $U_M=0$.
The system parameters are same as Fig. \ref{fig1}.
}\label{fig4}
\end{figure}

\begin{figure*}
\centering
\includegraphics[width=1.5\columnwidth]{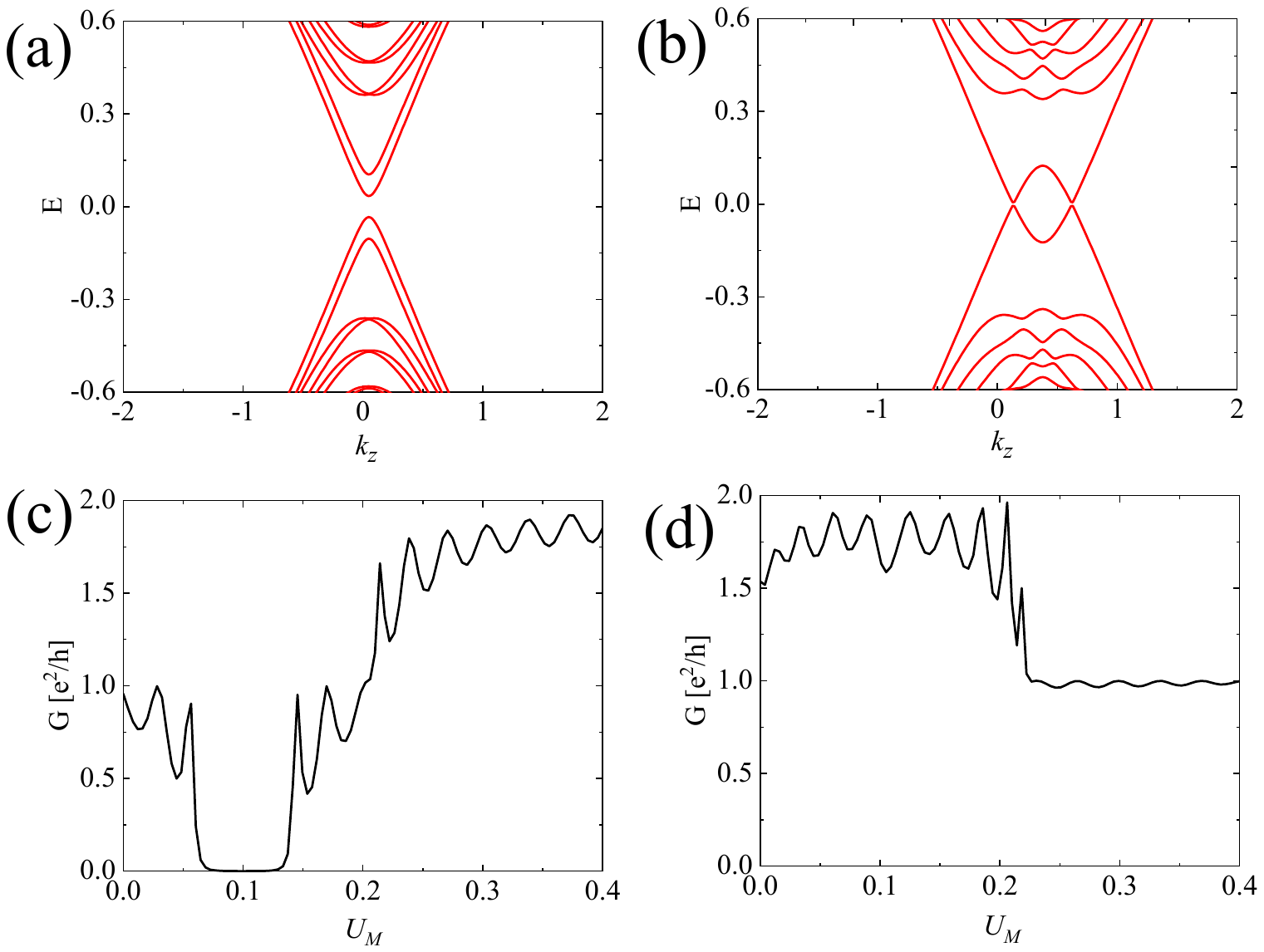}\\
\caption{The bands structures of SOTI with magnetic flux $\phi=0.004, 0.03$
are shown in (a) and (b). The transmission probability as a function of potential energy $U_M$
with constriction thickness $L_y=4a$ are shown in (c) and (d).
The system parameters are same as Fig. \ref{fig1}.
}\label{fig5}
\end{figure*}

We calculate the two terminal conductance in 3D SOTIs constrictions.
According to the Landauer - B$\ddot{\text{u}}$ttiker formula,
the conductance is
\begin{equation}\label{G}
  G=\frac{e^2}{h}T_{ij},
\end{equation}
where $T_{ij}$ is the transmission probability from lead $j$
to lead $i$. We use quantum transport package, Kwant \cite{groth2014kwant}, to simulate
the 3D SOTIs constrictions and calculate the transmission probability $T_{ij}$.
In order to run numerical calculations, we write the model in Eq. ($\ref{Hk}$) on a cubic lattice as
\begin{equation}\label{Hr}
\begin{aligned}
  H&=\sum_i{c_i^{\dagger} M\sigma_0\tau_z c_i}\\
  &+ \Bigg \{ \sum_i{{c_{i+x}^{\dagger} \bigg [ \frac{1}{2}(\Delta_2\sigma_0\tau_y + t\sigma_0\tau_z + i\Delta_1\sigma_x\tau_x) \bigg ] c_i}}\\
  &- \sum_i{c_{i+y}^{\dagger} \bigg [ \frac{1}{2}(\Delta_2\sigma_0\tau_y + t\sigma_0\tau_z + i\Delta_1\sigma_y\tau_x) \bigg ] c_i}\\
  &+ \sum_i{c_{i+z}^{\dagger} \bigg [ \frac{1}{2}(t\sigma_0\tau_z+i\Delta_1\sigma_z\tau_x) \bigg ] c_i} + h.c. \Bigg \},
\end{aligned}
\end{equation}
where $c_i=(c_{a,\uparrow,i},c_{b,\uparrow,i},c_{a,\downarrow,i},c_{b,\downarrow,i})$
are the annihilate operators of electron with spin up and spin down in $a$ and $b$ orbits at site $i$.

In Fig. \ref{fig3}, we plot two terminal conductance as a function of potential energy of constrictions region $U_M$
with incident energy $ie=0.1$ for $L_y=8a$ (black dot line), $L_y=6a$ (red dashed line) and $L_y=4a$ (blue solid line), respectively.
As one can see, the conductance has Fabry-Perot oscillation pattern. For $L_y=4a$,
there is a zero conductance plateau when increasing potential energy $U_M$.
Since we increase the potential energy $U_M$, the energy bands would shift up.
When the Fermi level fall in the energy gap of constrictions region,
the central region can be consider as a barrier. With broadening the thickness $L_y$,
the finite size effect weakening which means the tunnelling probability between opposite
hinges will weaken. The electrons from one lead propagate to another lead become possible,
so that the zero conductance plateau disappear for $L_y=8a$ and $L_y=6a$.

In addition, we also plot the conductance as a function of length of central region
as shown in Fig. \ref{fig4} with incident energy $ie=0.1$, potential energy $U_M=0$ for thickness
$L_y=8a$ (black dot line), $L_y=6a$ (dashed red line) and $L_y=4a$ (solid blue line), respectively.
As expected, the Fabry-Perot oscillation pattern appears when incident energy do not
fall in energy gap of central region. Moreover, the oscillation amplitude decreases with
increasing thickness $L_y$. When increasing the thickness $L_y$, the hybridization
of hinge states will weaken, giving rise to weakened backscattering probability.
Therefore, the Fabry-Perot amplitude decreases and the maximum value of the oscillation approaches the number of transport channels.

The key reason of Fabry-Perot oscillation is multiple backscattering
which results from the coupling of chiral hinge states.
The motion of the electrons will be modulated by magnetic field, so that
Fabry-Perot oscillation pattern would be changed with different magnetic field.
We consider that the $x$ direction magnetic field only imposes at constrictions,
$B=(B,0,0)$. Here, we only consider the orbital effect without Zeeman effect.
Hence, the hopping in Hamiltonian $H$ Eq. (\ref{Hr}) acquires the Peiels phase factor $e^{i\phi_{ij}}$,
\begin{equation}
  \phi_{ij}=\frac{e}{\hbar}\int_{r_i}^{r_j} A(r)\cdot dr,
\end{equation}
where $A(r)$ is the vector potential, which is given by $B=\nabla \times A$.
Here, we choose the Landau gauge $A=(0,0,By)$, thus only $z-$direction hopping should acquire the Peiels phase factor.
The magnetic field strength is characterized by the flux per unit cell,
\begin{equation}
  B=\frac{\phi\Phi_0}{a^2},
\end{equation}
where $\Phi_0$ is the quantum flux.

\begin{figure}
\centering
\includegraphics[width=\columnwidth]{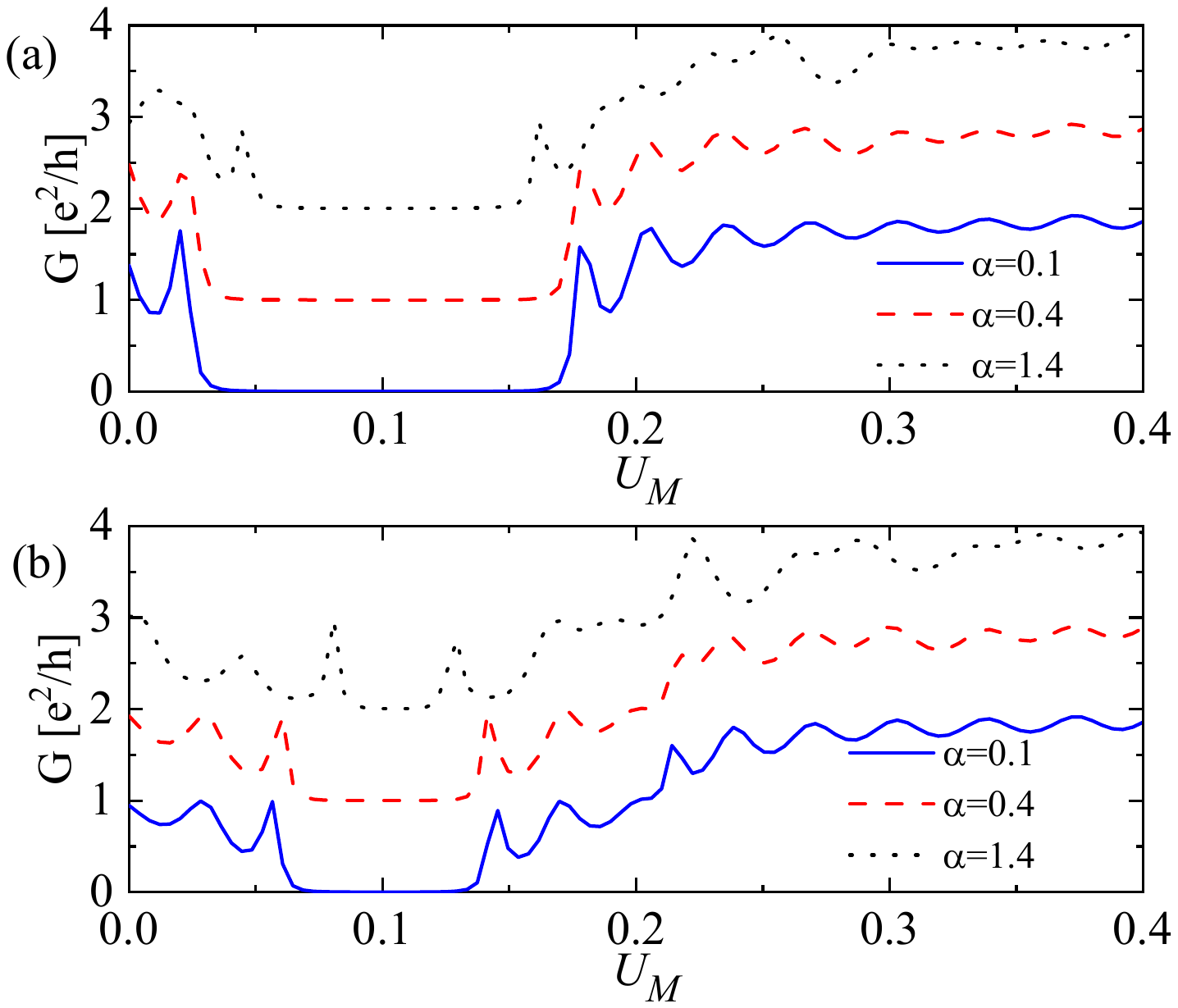}\\
\caption{The transmission probability as a function of potential energy $U_M$
with magnetic flux is shown in (a) $\phi=0$ and (b) $\phi=0.004$
with different disorder strengths $\alpha=0.1, 0.4, 1.4$.
An offset 1 is imposed to adjacent curves for clarity.
The system parameters are same as Fig. \ref{fig1}.
}\label{fig6}
\end{figure}

In Fig. \ref{fig5} (a) and (b), we plot energy spectrum of the constrictions
with the thickness $L_y=4a$ and the flux $\phi$ is $0.004$ and $0.03$, respectively.
One can see that the energy gap resulting from the finite size effects decreases
and even closed. Therefore, the zero conductance plateau will decrease and even disappear.
Moreover, the degenerate bands is split two subbands so that the electrons can
only localized at two hinges of sample when incident energy falls between two split subbands.
It means there must be a new oscillation pattern in two terminal conductance.
In Fig. \ref{fig5} (c) and (d), we plot the conductance as function of potential energy $U_M$
with different magnetic flux $\phi=0.004$ and $\phi=0.03$, respectively.
As expected, one can see the decreased or even disappeared zero conductance plateau
and the new oscillation conductance.

In the real materials, disorder is always unavoidable. Therefore, we should
confirm whether the disorder affect the interference pattern.
In Fig. \ref{fig6}, we show the numerical results for the disorder distributed
in constrictions with different strength, curves are offset by
1 for clarity. We can see the conductance of
Fabry-Perot oscillation is robust to moderate disorder. We expect these oscillation
pattern will be observed in experiments.

\section{Summary and Outlook}\label{so}
To summarize, we have investigated Fabry-Perot interference in 3D SOTIs constrictions.
Due to the coupling hinge states, there are the multiple reflection process in our setup.
The Fabry-Perot interference can be observed by measuring the two terminal conductance
as changing potential energy $U_M$ or length $L_M$. We also show the impact of
the magnetic field on the interference patterns. Finally, we show our numerical
results is robust to moderate disorder. In this paper, although, we choose the 3D SOTIs with chiral hinge
states as our model, the Fabry-Perot interference will be also observed in
the 3D SOTIs with helical hinge states. Furthermore, the constriction setup of helical hinge states
may have a potential application in spintronics.

\begin{acknowledgments}
This work was supported by the National
Natural Science Foundation of
China under Grant No. 12304063 and the startup
grant at Hubei Normal University.
\end{acknowledgments}

%

\end{document}